\documentclass[aps,prb,twocolumn,groupedaddress,showpacs,showkeys]{revtex4}
\usepackage[dvips]{graphicx,color}

\newcommand{\nal}{Ni$_3$Al}

\newcommand{\fatr}{\mathbf{r}}

\newcommand{\fatq}{\mathbf{q}}

\newcommand{\fatm}{\mathbf{m}}
\newcommand{\fatM}{\mathbf{M}}
\newcommand{\fatH}{\mathbf{H}}

\newcommand{\micron}{$\mu$m}

\newcommand{\pgnfigure}[3]{\begin{figure}\includegraphics[height=#1cm,clip=true]
{#2.eps}\caption{\label{#2}#3}\end{figure}}
\newcommand{\pgnfigures}[5]{\begin{figure}\includegraphics[height=#1cm,clip=true]
{#2.eps} \par \par \includegraphics[height=#3cm,clip=true]{#4.eps}
\caption{\label{#2}#5}\end{figure}}
\newcommand{\pgnformula}[2]{\begin{equation} \label{#1} #2 \end{equation}}

\begin{document}


\title{Spin-fluctuation dominated electrical transport of Ni$_3$Al at high pressure}


\author{P. G. Niklowitz}
\email[e-mail: ]{niklowit@drfmc.ceng.cea.fr}
\altaffiliation[present address: ]{D\'{e}partement de Recherche
Fondamentale sur la Mati\`{e}re Condens\'{e}e, SPSMS, CEA
Grenoble, 38054 Grenoble Cedex 9, France}
\author{F. Beckers}
\author{G. G. Lonzarich}
\affiliation{Cavendish Laboratory, University of Cambridge,
Madingley Road, Cambridge CB3 0HE, UK}
\author{G. Knebel}
\author{B. Salce}
\author{J. Thomasson}
\author{N. Bernhoeft}
\author{D. Braithwaite}
\author{J. Flouquet}
\affiliation{D\'{e}partement de Recherche Fondamentale sur la
Mati\`{e}re Condens\'{e}e, SPSMS,CEA Grenoble, 38054 Grenoble
Cedex 9, France}


\date{\today}

\begin{abstract}
We present the first study of a magnetic quantum phase transition
in the itinerant-electron ferromagnet Ni$_3$Al at high pressures.
Electrical resistivity measurements in a diamond anvil cell at
hydrostatic pressures up to 100~kbar and temperatures as low as
50~mK indicate that the Curie temperature collapses towards
absolute zero at a critical pressure $p_c=82 \pm 2$~kbar. Over
wide ranges in pressure and temperature, both in the ferromagnetic
and paramagnetic states, the temperature variation of the
resistivity is found to deviate from the conventional Fermi-liquid
form. We consider the extent to which this deviation can be
understood in terms of a mean-field model of enhanced spin
fluctuations on the border of ferromagnetism in three dimensions.
\end{abstract}


\maketitle

\section{Introduction}

The electronic properties of metals on the border of magnetic
phase transitions at low temperatures are often found to exhibit
temperature dependences at variance with the predictions of the
standard model of a normal Fermi liquid.  Early attempts to
explain such non-Fermi-liquid behaviour have been based on a
mean-field treatment of the effects of strongly-enhanced
spin-fluctuations (paramagnons). \cite{her76a,mil93a,mor85a}

For a metal on the border of ferromagnetism in three spatial
dimensions (3D), this mean-field spin-fluctuation model predicts a
$T^{5/3}$ temperature dependence of the resistivity,\cite{mat68a}\
instead of the conventional $T^2$ temperature dependence of a
normal metal at low temperatures. The $T^{5/3}$ variation of the
resistivity is a consequence of an underlying quasiparticle
scattering rate that varies linearly with the excitation energy
$E$ of a quasiparticle near the Fermi level. This is the behaviour
associated not with a Fermi liquid, for which the quasiparticle
scattering rate varies as $E^2$, but of a cross-over state known
as the marginal Fermi liquid.\cite{var89a,hol73a,bay91a,nik04a}\ A
review of several underlying models that yield a marginal
Fermi-liquid form of the electron self energy is given in
Ref.~\onlinecite{bay91a}.

In contrast to a phenomenological model introduced to describe the
normal state of the cuprates,\cite{var89a}\ the marginal Fermi
liquid that arises on the border of an itinerant-electron
ferromagnet in 3D is due to the scattering of long-wavelength spin
fluctuations that are relatively ineffective in reducing the
current. This leads to a temperature dependence of the transport
relaxation rate or resistivity that is characterized by an
exponent above unity, i.e., 5/3, but still below that expected for
a conventional Fermi liquid.\cite{mat68a}

In this paper we re-examine the behaviour of the resistivity on
the border of itinerant-electron ferromagnetism in the relatively
simple case of \nal\ in which the cross-over from ferromagnetism
to paramagnetism is achieved by the application of hydrostatic
pressure.  \nal\ can be prepared in a pure stoichiometric form and
crystallizes in a simple cubic (Cu$_3$Au) structure.\cite{mof83a}\
At ambient pressure it orders ferromagnetically below 42~K with a
small average moment of 0.075~$\mu_B$/Ni in the limit of low
temperature and low magnetic
field.\cite{flu69a,boe69a,sas84a,ber82a}\

For a test of the predictions of the mean-field spin-fluctuation
model, \nal\ appears to offer advantages over a number of other
metals on the border of magnetism. The magnetic critical point can
be tuned via hydrostatic pressure as opposed to chemical doping
(as in, e.g., Pd(Ni))\cite{nic99a}\ that can introduce new physics
of disorder not incorporated fully in the mean-field
spin-fluctuation model. In contrast to other d-metals (e.g., MnSi,
CoSi$_2$),\cite{pfl97a,the97a,mus02a,bar04a}\ \nal\ has a more
nearly continuous quantum phase transition and (unlike MnSi) does
not exhibit a spin-spiral structure due to the
Moriya-Dsyaloshinski interaction arising from lack of space
inversion symmetry.  Also, in contrast to the nearly magnetic
f-metals (e.g., CeCu$_{6-x}$Au$_x$, CePd$_2$Si$_2$,
YbRh$_2$Si$_2$, UGe$_2$),\cite{sch00a,mat98a,cus03a,sax00a} the
spin-orbit interaction in \nal\ is relatively weak, the energy
bands are relatively broad and the spin fluctuations are not local
in real space, features that may be necessary for the
applicability of the mean-field spin-fluctuation model in its
present form.

As a reference system for the study of the border of metallic
ferromagnetism, \nal\ is also convenient because, along with its
close relative, the nearly ferromagnetic metal Ni$_3$Ga, it has
been extensively studied using a wide range of experimental
techniques,\cite{flu73a,bui81a,sig84a,bui83a,ber83a,ber86a,ber89a,dha89a,sem00a,kil03a}\
all of which indicate the importance of spin fluctuations in these
materials.\cite{agu04a,maz04a}\ The combined results in these two
systems define the parameters of the mean-field spin-fluctuation
model that may be relevant to interpreting the behaviour of \nal\
from its low-pressure ferromagnetic phase to the high pressure
paramagnetic phase near and above $p_c$ where a marginal Fermi
liquid cross-over state may be expected to be
observed.\cite{lon85a}\

\section{Experimental details}

\nal\ samples were pressurised using a diamond anvil cell (DAC)
described in detail elsewhere.\cite{tho91a}\ The cell employed
anvils with culet diameter of 1~mm, a stainless-steel gasket with
central hole for the sample and several ruby chips, and Argon as
the pressure medium. After compression, the sample space was
typically 400~\micron\ in diameter and 50~\micron\ in thickness.
The pressure was determined from the fluorescence of the tiny ruby
chips in the sample space.\cite{jay83a}\ The noble gas pressure
medium ensured that the applied pressure was very hydrostatic. The
pressure inhomogeneity over the sample was only 3\%\ of the
applied pressure.

The samples were prepared from the melt by radio frequency heating
starting with zone refined Ni and Al having residual resistivity
ratios of over 2000. High homogeneity samples were obtained by
suitable stirring of the high purity stoichiometric melt, followed
by rapid quenching and then annealing for up to 6 days. Small
single crystals were spark cut from the resulting ingot and then
characterised by means of microprobe analysis, transmission
electron microscopy, Laue x-ray diffraction, mass spectroscopy,
magnetic hysteresis and resistivity ratio measurements. These
studies did not reveal evidence of precipitates of other phases or
of total metallic impurity levels in excess of 20 ppm. The
specimens selected for pressure studies in a $^4$He system and in
a dilution refrigerator had residual resistivity ratios of 27 and
29, respectively, and had been used previously for the study of
the de Haas-van Alphen (dHvA) effect.\cite{sig84a}\ Tiny samples
of the size required to fit into the DAC were cut to an initial
thickness of 100~\micron\ by means of low-power spark erosion and
thinned to the final thickness of 10~\micron\ by chemical
polishing.

The resistivity was measured by a sensitive ac 4-terminal
technique. Four 12~\micron\ gold wires were attached to the sample
by a micro-spot-welding technique that gave low contact resistance
and thus low excitation-current heating and high detection
sensitivity. Damage to the sample was minimized by spot welding
with very low power immediately after the surface of the sample
had been cleaned by chemical polishing. The gold wires were passed
between one diamond anvil and one side of the stainless steel
gasket insulated with a layer of 1266 Stycast epoxy mixed with a
saturated concentration of Al$_2$O$_3$ powder.\cite{tho98a}\ The
resistivity was measured in a $^4$He system from 1.5 to 40~K with
an excitation current of 1~mA, and in a dilution refrigerator from
50~mK to 6.5 K with an excitation current of 100~mA. The dilution
refrigerator contained a low temperature transformer and both
refrigerators were equipped with a variable force application
mechanism that allowed the in-situ changes of the pressure in the
DAC.\cite{sal00a}

\section{Results}

\subsection{Magnetic phase diagram of \nal}

\pgnfigure{13}{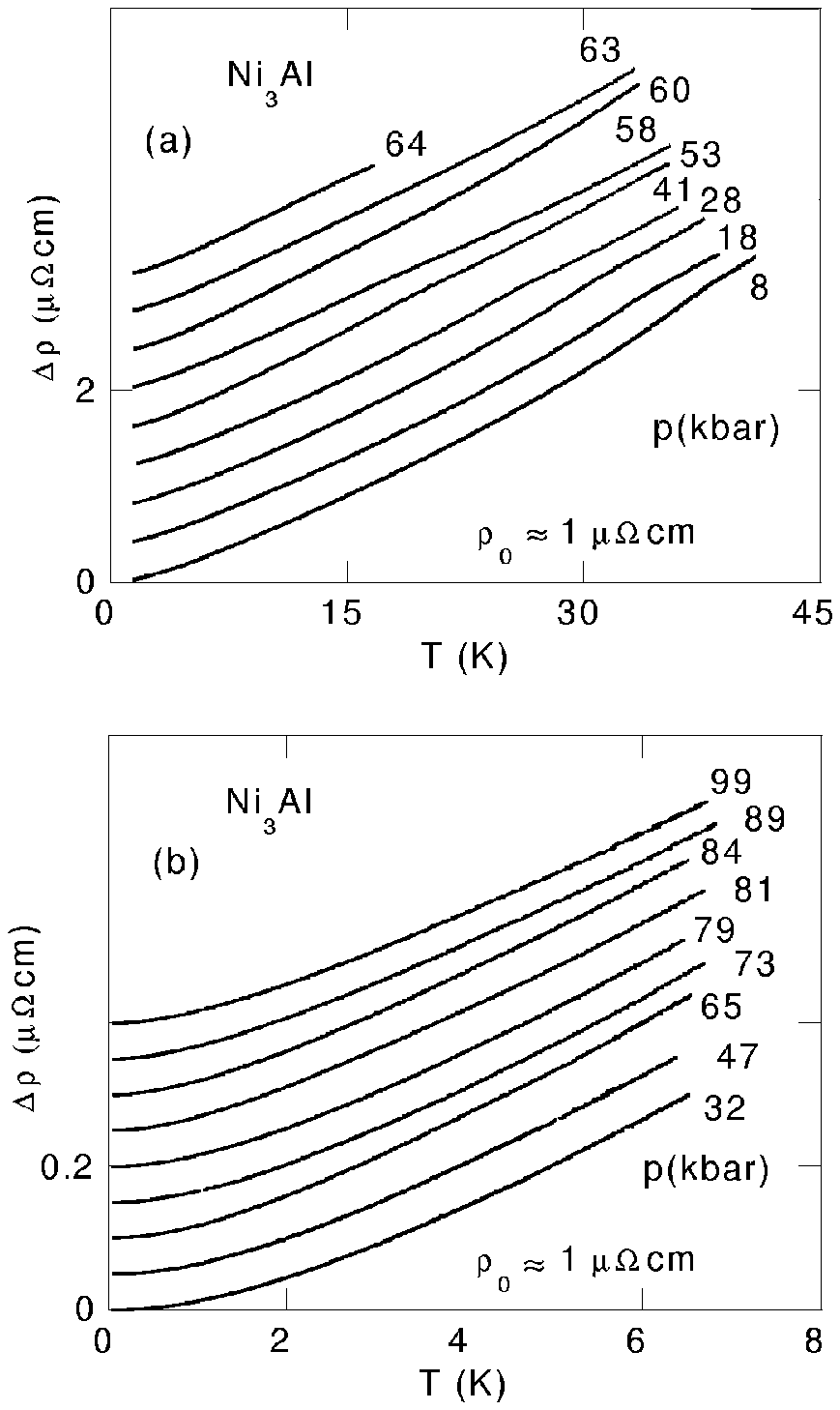}{The temperature-dependent part of the
electrical resistivity $\Delta\rho=\rho-\rho_0$ at various
pressures. Measurements were carried out in a $^4$He system (a)
and a dilution refrigerator (b). Upon pressure application
$\rho_0$ increased irreversibly by about 1~$\mu\Omega$cm and thus
the intrinsic pressure dependence of $\rho_0$ could not be
inferred. The curves have been shifted vertically for clarity.}

\pgnfigure{6}{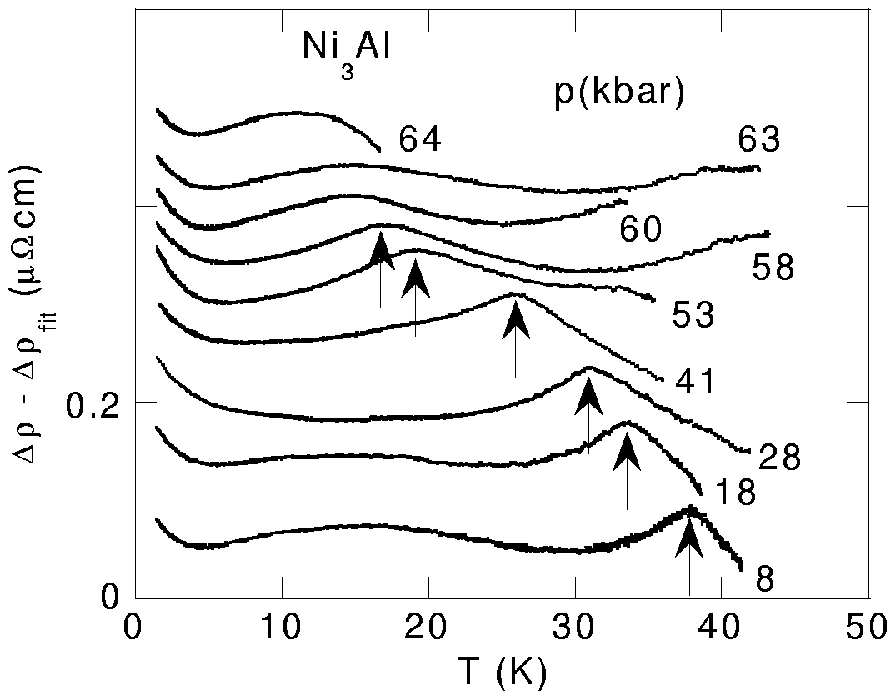}{Signature of the ferromagnetic transition
of \nal\ in the temperature dependence of the resistivity. The
ferromagnetic transition at $T_{Curie}$ indicated by an arrow
shows up in a plot of $\Delta\rho-\Delta\rho_{fit}$ where
$\Delta\rho_{fit}$ is a smooth second-order polynomial fit of the
data over the entire experimental range. The same values of
$T_{Curie}$ are obtained within experimental error from plots of
$\partial \rho/\partial T$ vs $T$. The curves have been shifted
vertically for clarity.}

\pgnfigure{6}{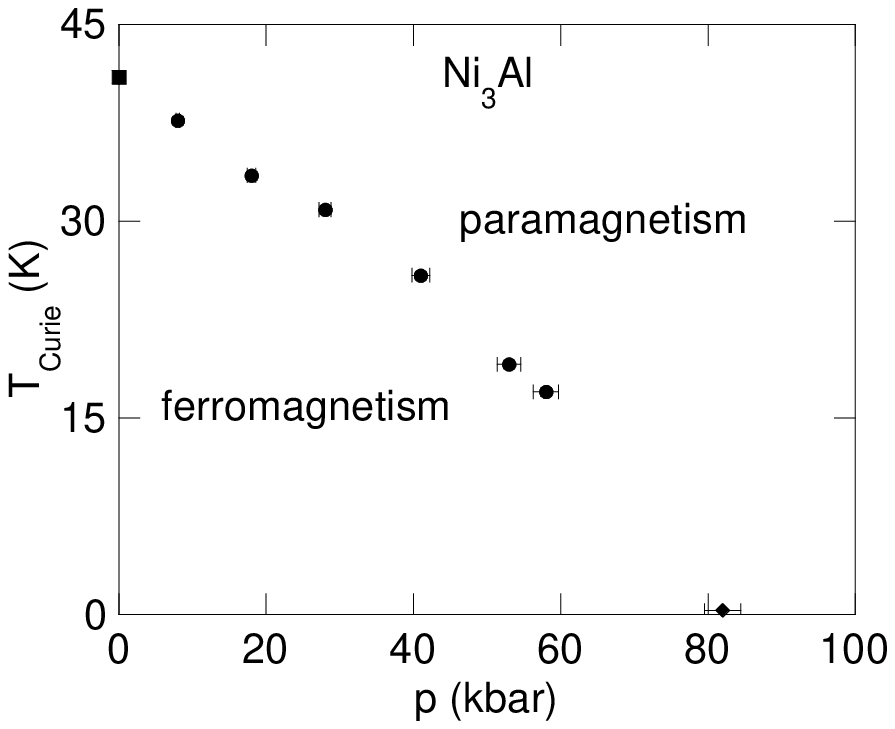}{The proposed magnetic
temperature-pressure phase diagram of \nal. The full circles
represent the peaks in the resistivity data
$\Delta\rho-\Delta\rho_{fit}$ in Figure~\ref{ppnalrtc}. The square
represents the ferromagnetic transition $T_{Curie}=41\pm1$~K at
ambient pressures inferred in samples similar to those studied
here by means of both resistivity and magnetic susceptibility
measurements. The diamond defines the critical pressure
$p_c=82\pm2$~kbar inferred from the peak in the $T^2$ coefficient
of the resistivity vs pressure plotted in the inset of
Figure~\ref{ppnaldlt}b.}

Our measurements of the temperature dependence of the resistivity
of \nal\ at different pressures are shown in
Figure~\ref{ppnaldata}.  The temperature dependent part of the
resistivity is defined by $\Delta\rho=\rho-\rho_0$, where $\rho_0$
is the residual resistivity inferred by a suitable extrapolation
of $\rho$ vs $T$ to absolute zero of temperature. The small values
of $\rho_0$, initially of the order of 1~$\mu\Omega$cm for our
samples, were found to increase irreversibly by about
1~$\mu\Omega$cm during the pressure experiments, and thus the
intrinsic pressure dependence of $\rho_0$ could not be inferred.
At low and intermediate pressures, the resistivity curves exhibit
very weak anomalies in $\rho$ vs $T$ centred at a temperature that
coincides at ambient pressure with the Curie temperature
$T_{Curie}$. It is natural to associate the pressure dependence of
these anomalies with the pressure dependence of $T_{Curie}$. The
anomalies can be made more evident by plotting the derivative
$\partial\rho/\partial T$ or by subtracting from $\rho$ a smooth
polynomial fit of the form $\Delta\rho_{fit}=c_0+c_1T+c_2T^2$. The
curves $\Delta\rho-\Delta\rho_{fit}$ shown in
Figure~\ref{ppnalrtc}\ clearly reveal a peak that collapses
monotonically with increasing pressure. Plots of
$\partial\rho/\partial T$ and of $\Delta\rho-(c_0+c_1T)$ also lead
to peak positions consistent with those in Figure~\ref{ppnalrtc}.
The weakening of the peaks with increasing pressure is consistent
with the predictions of the mean-field spin-fluctuation model. We
note that near $T_{Curie}$ the magnetic contribution to $\partial
\rho/\partial T$ is quite generally expected to be proportional to
the magnetic contribution to the heat
capacity,\cite{fis68a,ale76a}\ which, in the mean-field
spin-fluctuation model, is predicted to decrease with decreasing
$T_{Curie}$.\cite{mur72a} Particularly at high pressures, however,
the peaks may also be reduced by broadening due to pressure
inhomogeneities. This effect grows with increasing downward
curvature of $T_{Curie}$ vs $p$ and may be expected to lead to a
disappearance of the Curie point anomaly in $\rho$ vs $T$ near
$p_c$.

The pressure dependence of $T_{Curie}$ inferred from the
resistivity anomalies is shown in Figure~\ref{ppnalrphdiag}.
$T_{Curie}$ collapses with increasing pressure with an initial
gradient of approximately $-0.4\pm 0.05$~Kkbar$^{-1}$, which is
close to the value reported in an early low-pressure study.
\cite{bui76a}\ Assuming a slightly stronger than linear variation,
$T_{Curie}$ vs $p$ extrapolates to a value close to $p_c=82\pm
2$~kbar defined by the peak in the low-temperature limit of
$A=\Delta\rho/T^2$ vs $p$ (see Figure~\ref{ppnaldlt}). However,
the lack of evidence of a clear signature of $T_{Curie}$ in
$\Delta\rho$ vs $T$ above 60~kbar leaves some doubt as to the true
value of the critical pressure at which $T_{Curie}$ vanishes.

We note that ferromagnetism in \nal\ can also be suppressed by Al
doping.\cite{boe69a,yos92a}\ However, the high sensitivity of
$T_{Curie}$ to dopant concentration $x$ in Ni$_{75-x}$Al$_{25+x}$
has made it difficult to produce a detailed temperature-dopant
phase diagram for comparison with our temperature-pressure phase
diagram. In previous work $x$ was varied in steps of 0.5,
corresponding to steps of the order of 100~kbar, and the critical
value of $x$ where $T_{Curie}$ vanishes is estimated by
interpolation of the data to be about 0.4. Doped specimens have
higher $\rho_0$ and thus lower values of $\Delta\rho/\rho_0$,
making measurements of the temperature dependence of $\rho$ more
difficult. Also, doped samples may have significant
inhomogeneities in $T_{Curie}$ and may involve physics not
included in the mean-field spin-fluctuation model.

\subsection{Temperature dependence of the resistivity}

The evolution with pressure of the temperature dependence of the
resistivity is shown in Figures~\ref{ppnaldrr0_1}\ and
\ref{ppnaldrr0_2}. The resitivity is plotted against $T$,
$T^{3/2}$, $T^{5/3}$, and $T^2$ for three pressures corresponding
to the lower and upper end of the pressure range of the data in
Figure~\ref{ppnaldata}b as well as the critical pressure.

Overall the temperature dependent part of the resistivity appears
to be only weakly pressure dependent over the entire pressure
range investigated up to nearly 100~kbar. This is in marked
contrast to the behaviour of MnSi and, in particular, of narrow
f-band materials that in general exhibit obvious and strong
variations of the resistivity upon crossing the critical pressure
over a pressure range of typically only a few kbar. \nal\ differs
in its low-temperature resistivity from typical f-band systems in
being essentially very near a magnetic quantum critical point over
a wide pressure range.

Figures~\ref{ppnaldrr0_1} and \ref{ppnaldrr0_2} also show that
even at low temperatures, 0.05~K to 7~K, the resistivity cannot be
described in terms of a simple power law. The overall best fit in
this temperature range is to an exponent of the order of 3/2.
Below 4~K, however, the best fit is to an exponent of the order of
5/3 and below 1~K to an exponent of about 2.  This behaviour is
highlighted in the temperature variation of the logarithmic
derivative of the resistivity $\partial\ln\Delta\rho/\partial\ln
T$ shown in Figure~\ref{ppnaldexpT}, which defines the temperature
dependent resistivity exponent $n(T)$. At still higher
temperatures (well above 10~K), the resistivity exponent
eventually drops towards unity (see Figure~\ref{ppnaldata}a).

\pgnfigure{13}{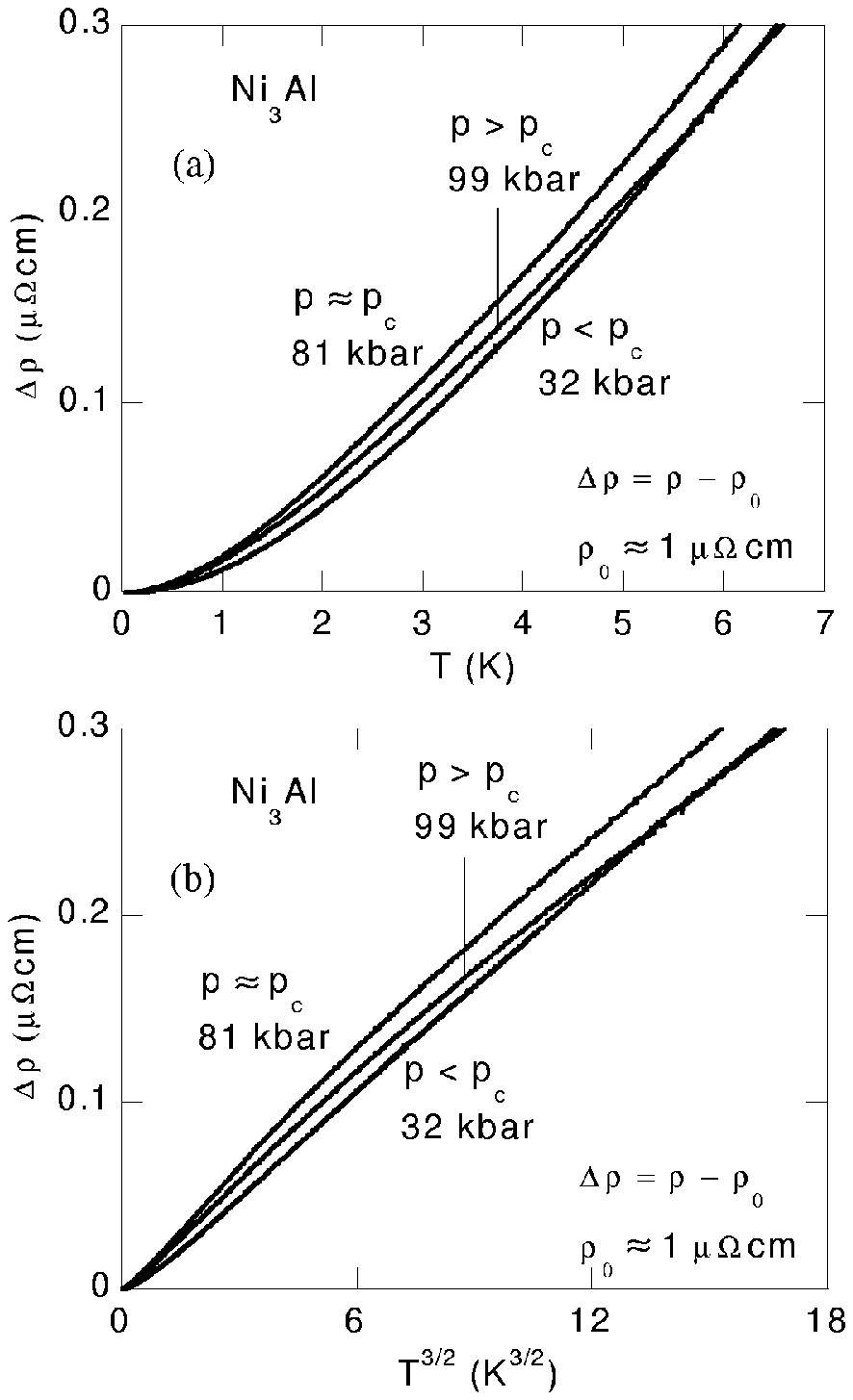}{Temperature-dependent part of the
resistivity $\Delta\rho$ of \nal\ versus $T$ (a) and versus
$T^{3/2}$ (b). $\Delta\rho$ varies more strongly than linear with
temperature.  At low temperatures, $\Delta\rho$ rises more
strongly than $T^{3/2}$ and at higher temperatures $\Delta\rho$
rises more weakly than $T^{3/2}$. As shown in
Figure~\ref{ppnaldexpT}\ the resistivity exponent, defined by the
logarithmic derivative of $\Delta\rho$, is not constant and varies
strongly with temperature even in the liquid helium temperature
range.}

\pgnfigure{13}{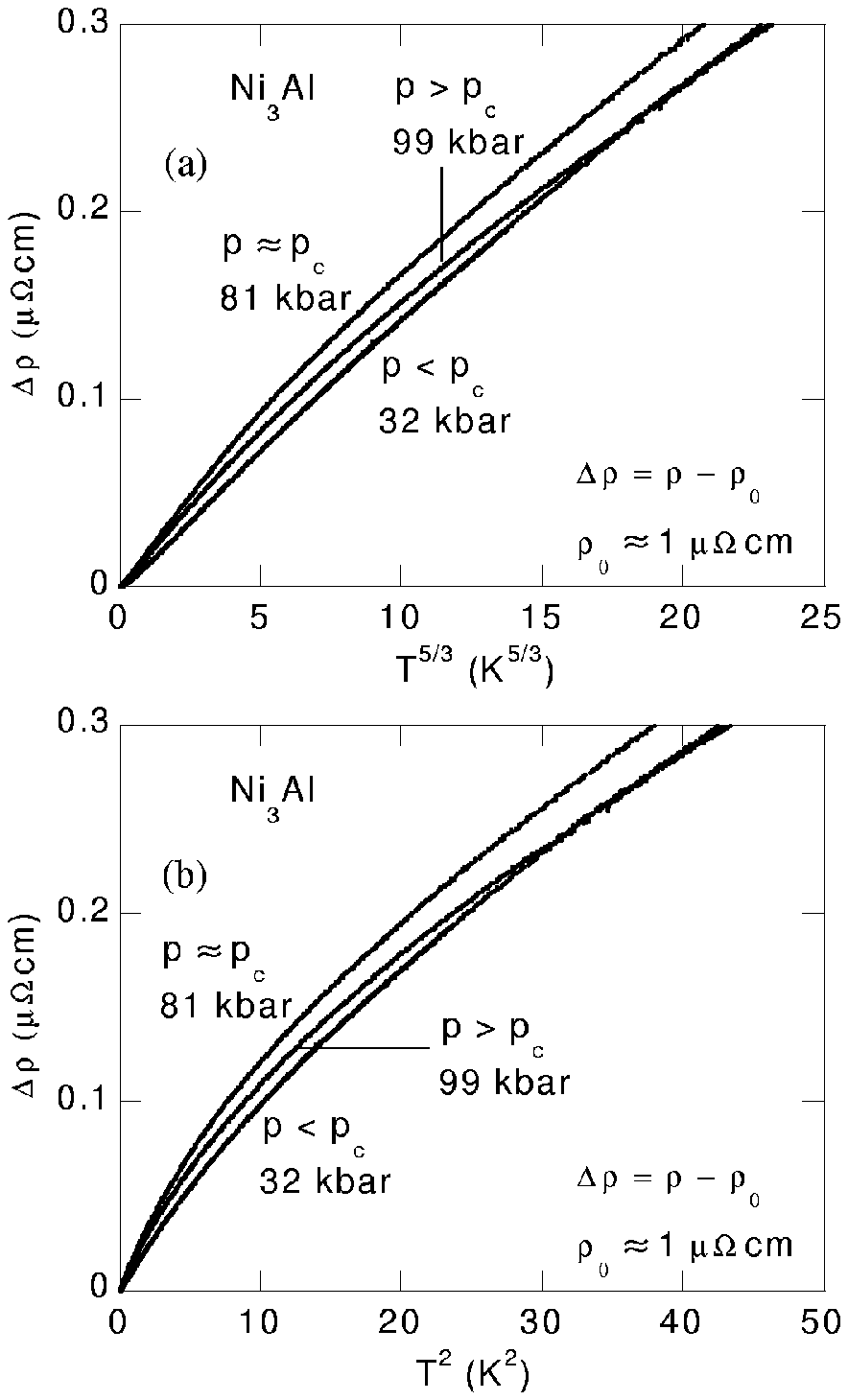}{Temperature-dependent part of the
resistivity $\Delta\rho$ of \nal\ versus $T^{5/3}$ (a) and versus
$T^2$ (b). The resistivity exponent defined by the logarithmic
derivative of $\Delta\rho$ is not constant and increases with
decreasing temperature from about 1.5 at 5~K towards 2 below 1~K
(Figure~\ref{ppnaldexpT}).}

\pgnfigure{6}{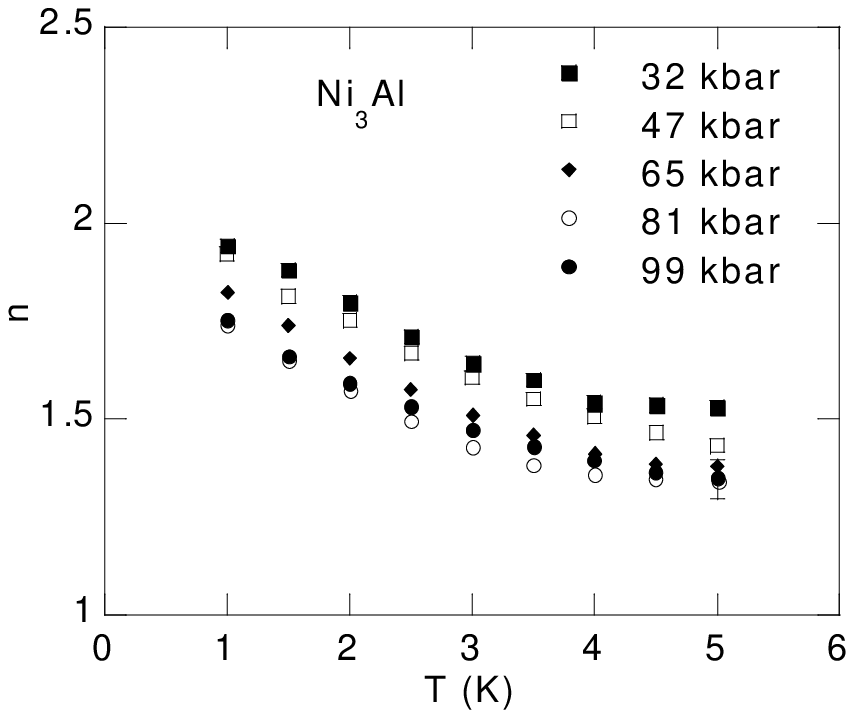}{The resistivity exponent of \nal\
defined by the logarithmic derivative of the resistivity,
$n=\partial ln\Delta\rho/\partial ln T$. The resistivity exponent
rises with decreasing temperature from about 1.5 at 5~K towards 2
below 1~K. At each $T$ the exponent is a minimum at around $p_c$.
We note that there in no extended temperature regime for either a
$T^2$ or $T^{5/3}$ behaviour. The error bar given at 5~K applies
to all data points shown in the plot.}

\pgnfigure{13}{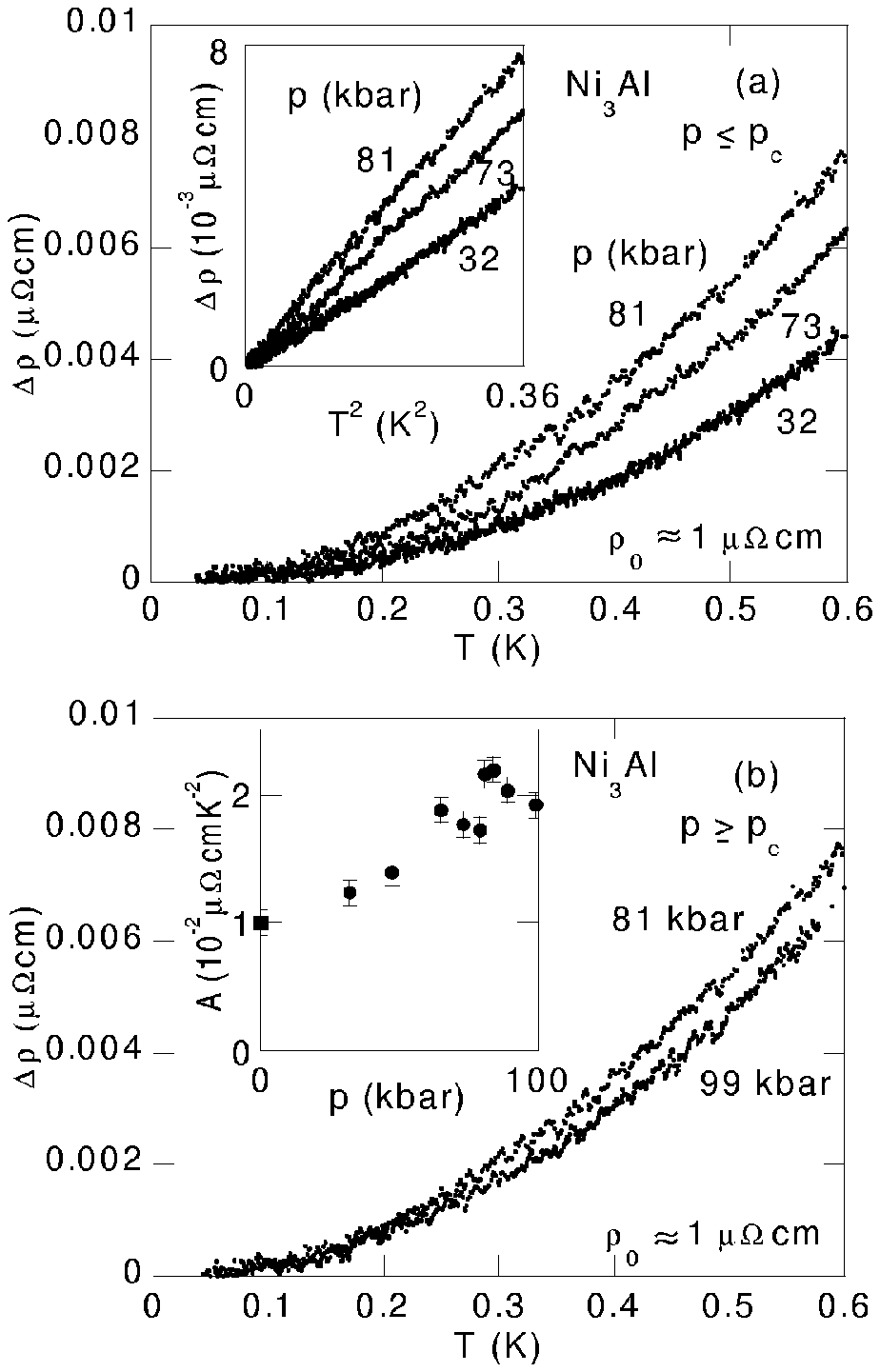}{Temperature dependent part of the
resistivity $\Delta\rho$ of \nal\ in the millikelvin range. In
this range where $\Delta\rho$ is only of the order of 1-2\%\ of
$\rho_0$, $\Delta\rho$ is approximately of the form $AT^2$ for our
sample. The $T^2$ coefficient $A$ increases with rising pressure
up to $p_c=82\pm2$~kbar (a) and falls again beyond $p_c$ (b). As
shown in the inset of Figure (b) the peaks value of $A$ at $p_c$
is a factor of 2 higher than the zero-pressure value of
0.01~$\mu\Omega$cmK$^{-2}$.\cite{sas84a}\ $A$ has been determined
from a fit of $\Delta\rho$ vs $T^2$ in the range 50~mK to 600~mK.}

\pgnfigure{6}{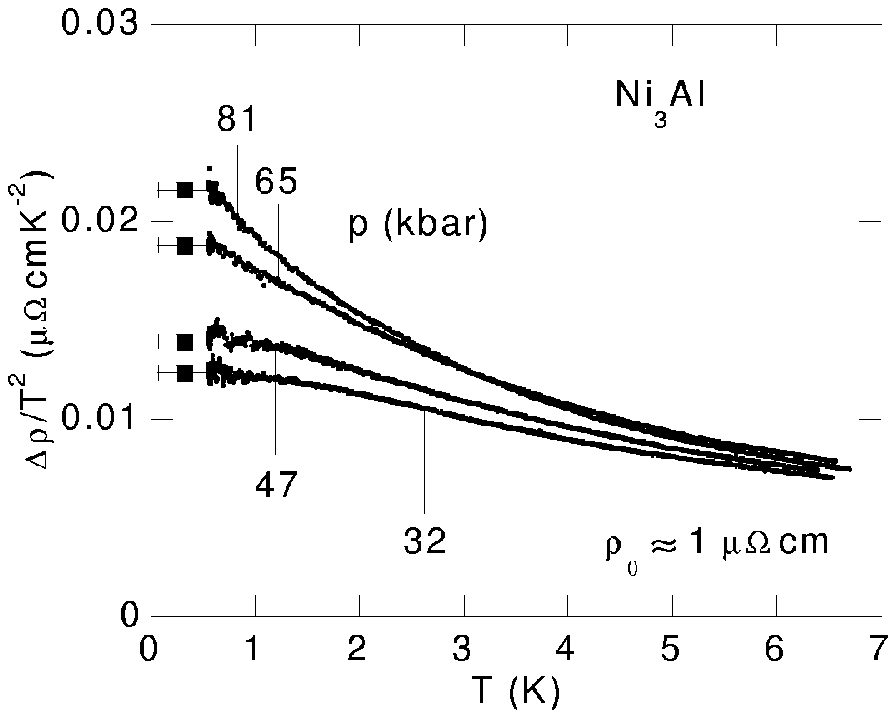}{Pressure dependence of $\Delta\rho/T^2$
versus $T$ in \nal. The ratio $\Delta\rho/T^2$ increases with
decreasing temperature and tends to saturate at low temperature.
The characteristic temperature at which saturation occurs
decreases with increasing pressure up to the critical pressure
($p_c=82\pm 2$~kbar). Correspondingly, the saturation value of
$\Delta\rho/T^2$ increases with pressure and peaks near $p_c$ (see
inset of Figure~\ref{ppnaldlt}b). We note that near $p_c$ the
saturation of $\Delta\rho/T^2$ is not evident except when
averaging over the data from 50 to 600~mK (squares) where
$\Delta\rho\ll\rho_0$.}

For low pressures where comparisons can be made our findings are
generally consistent with previous resistivity measurements in
stoichiometric samples \cite{flu69a,flu73a,sas84a,yos92a,ste03a}\
and in samples in which the ratio of the Ni and Al concentrations
were varied so as to suppress ferromagnetic order.\cite{flu73a}\
In all cases the temperature dependence of the resistivity is
found to deviate from the Fermi liquid form and to be described in
first approximation in terms of a resistivity exponent around 3/2
in the temperature range 1-30~K.  However, in contrast to the
present work, these earlier studies did not yield clear evidence
for a limiting $T^2$ form of the resistivity at very low
temperatures below 1~K.  In particular, Fluitman et al., found a
resistivity exponent of the order of 3/2 down to 200~mK in samples
with Ni to Al ratios tuned to the critical value where $T_{Curie}$
vanishes.\cite{flu70a}\ Furthermore, in a study of several
stoichiometric \nal\ samples, Steiner et al., found that while
above 1~K the resistivity exponent is sample independent, and
consistent with our work, below 1~K it varied significantly from
sample to sample, exhibiting in some cases weak upturns or
downturns below a few hundred mK.\cite{ste03a}\

We note that in all samples investigated thus far $\Delta\rho$ is
very small compared with $\rho_0$ at dilution refrigerator
temperatures (i.e., $\Delta\rho/\rho_0<2\%$ for $T<1$~K). Thus,
the behaviour of $\Delta\rho$ in this range may be sensitive to
sample heterogeneities and the precise way by which the samples
are prepared.  Thus, the form of the resistivity that we observe
below 1~K (Figure~\ref{ppnaldlt}) is not necessarily the property
of the ideal stoichiometric state of \nal. It is interesting to
note, however, that the $T^2$ coefficient $A$ of $\Delta\rho$ in
the mK range shows a peak (the peak position we define to be
$p_c$; inset of Figure~\ref{ppnaldlt}b). Plots of $\Delta\rho/T^2$
vs $T$ at different pressures in Figure~\ref{ppnaldsqdev}\ show
that the effect of pressure and the approach of a quantum critical
point near $p_c$ become most clearly evident in the low
temperature limit. We see that not only is the maximum of
$\Delta\rho/T^2$ vs $T$ at $p_c$ strongest at low temperatures but
also the crossover temperature above which $\rho$ clearly deviates
from the Fermi-liquid form (e.g., the temperature where
$\Delta\rho/T^2$ falls by 20\%\ of its value in the zero
temperature limit) reaches a minimum at this same pressure. At
81~kbar $\Delta\rho/T^2$ vs $T$ continues to grow with decreasing
temperature down to at least below 600~mK where
$\Delta\rho\ll\rho_0$ and a well-defined finite value of
$A=\Delta\rho/T^2$ as given in the inset of Figure~\ref{ppnaldlt}b
is obtained only after averaging over 500~mK. Although Fluitman et
al. did not report observing a $T^2$ form of $\Delta\rho$, they
nevertheless considered the behaviour of the average value between
1.2 and 4.2~K of the ratio $\Delta\rho/T^2$.\cite{flu73a}\ They
find that this ratio varied by a factor of about 1.3 in going from
a stoichiometric \nal\ sample to a non-stoichiometric sample with
critical doping where $T_{Curie}$ vanishes. This agrees well with
the variation versus pressure of $\Delta\rho/T^2$ in the same
temperature range (Figure~\ref{ppnaldsqdev}).

\section{Discussion}

In the following we discuss our experimental results in terms of
an elementary form of the mean-field spin-fluctuation model
discussed in the introduction.\cite{lon97a}\ Near the critical
pressure where ferromagnetic order in 3D vanishes, this model
leads to an electron self energy of a form at low energy
characteristic of a marginal Fermi liquid.\cite{dzy76a}\

We consider a 3D isotropic itinerant-electron system in which the
total spin is conserved and only long wavelength fluctuations of
the magnetization are strongly enhanced by an exchange molecular
field arising from the effects of the Coulomb interaction and the
Pauli principle. These conditions may be approximately satisfied
in the case of \nal\ for the reasons given in the introduction.
The essential features of the model are as follows.  We assume
that in the low $T$ limit the magnetization $\fatM(\fatr)$ in a
weak applied magnetic field $\fatH(\fatr)$ is given by a
Ginzburg-Landau equation of the form
\pgnformula{feos}{\fatH=a\fatM+b\fatM^3-c\nabla^2\fatM} where
$\fatM^3=(\fatM\cdot\fatM)\fatM$ and $b$ and $c$ are positive
constants. The latter conditions imply that the system undergoes a
continuous ferromagnetic transition when $a$ crosses zero. As in
the Landau mean-field model we assume $a$ is linear in $(p-p_c)$
for $p$ close to the critical pressure $p_c$, i.e., that
$a=\alpha(p- p_c)$, where $\alpha$ is a positive constant. An
analytic expansion of $a$ and the mean-field analysis given below
is thought to be plausible because the effective dimension
relevant to quantum phenomena in the $T\rightarrow 0$ limit is
greater than the upper critical dimension in
\nal.\cite{her76a,mil93a,mor85a}\ We also assume that relaxation
of a fluctuation of the magnetization $\fatM(\fatr,t)$ to
equilibrium is governed by Landau damping. Thus, a Fourier
component $\fatM_{\fatq}(t)$ of $\fatM(\fatr,t)$ decays
exponentially to the value given by Equation~\ref{feos} via a
relaxation function that is proportional to
$q=\left|\fatq\right|$. In zero applied field and small
$\fatM_{\fatq}$ in the paramagnetic state this implies
\pgnformula{frelax}{\frac{\partial\fatM_{\fatq}}{\partial t
}=-\gamma q(a+cq^2)\fatM_{\fatq}} where $\gamma$ is a positive
constant. We note that the coefficient of $(-\fatM_{\fatq})$ on
the right hand side of Equation~\ref{frelax} defines the $\fatq$
dependent relaxation rate $\Gamma_{\fatq}$ of magnetic
fluctuations. Thus, $\Gamma_{\fatq}$ is given by $\gamma q$ times
the inverse static susceptibility, which is defined as the linear
coefficient of $\fatM_{\fatq}$ in the Fourier transform of
Equation~\ref{feos}. Note that when $a\rightarrow 0$ the
relaxation rate becomes cubic in $q$, i.e., the dynamical exponent
$z$ is here equal to 3. Thus for $D=3$ the effective dimension
$D+z$ relevant to quantum critical phenomena is equal to 6, which
is indeed above the upper critical dimension of 4 for our
Ginzburg-Landau model.

The model defined by Equations~\ref{feos}\ and \ref{frelax}\ is
found to be consistent with ambient-pressure bulk magnetization
measurements\cite{boe69a}\ that yield the parameters $a$ and $b$
and ambient-pressure inelastic neutron scattering
measurements\cite{ber82b}\ that yield estimates of $c$ and
$\gamma$ in \nal. The values of these essentially ground state
parameters define the starting point of a model for the
temperature dependence of the magnetic equation of state, heat
capacity and resistivity.

The temperature dependence of the magnetic equation of state is
assumed to arise primarily from the effects of strongly enhanced
long-wavelength spin fluctuations rather than from incoherent
particle-hole excitations of conventional band theory.
Specifically, any temperature dependence is assumed to arise in a
self-consistent way from Equations~\ref{feos}\ and \ref{frelax}\
together with the Bose function $n_{\omega}$, which governs the
excitations of those spin fluctuations that are strongly excited
on the border of magnetic long-range order. Thus, the finite
temperature properties are consequences of a set of $T=0$
parameters and a universal thermal factor in analogy to the Debye
model for lattice vibrations or the Fermi liquid model for the
quasiparticle excitations of a normal metal at low temperatures.
The principal difference is that in the spin-fluctuation model the
relevant modes that are thermally excited are not normal modes in
the usual sense, but are characterized by a relaxation spectrum
(see Equation~\ref{frelax}). The self-consistent spin-fluctuation
model and the earlier non-self-consistent paramagnon model can
thus be viewed as being elementary examples involving the
statistical mechanics of open or dissipative
systems.\cite{lon86a,pip87a,edw92a}

\subsection{Magnetic phase boundary}

We consider first the effect of thermal fluctuations of the
magnetization on the magnetic equation of state
(Equation~\ref{feos}). These fluctuations may be imagined to arise
from the effect of a random field of zero mean added to the left
hand side of Equation~\ref{feos}\ which then drives a random
magnetization of zero mean added to the average magnetization in
each term on the right hand side of Equation~\ref{feos}. In the
absence of the non-linear term on the right hand side of
Equation~\ref{feos} such fluctuations when averaged over an
ensemble yield no effect. However, the non-linear term
($b\fatM^3$) leads to corrections that depend on the variance of
the local magnetization. In lowest order, in the paramagnetic
state, for example, one finds that the linear coefficient in
Equation~\ref{feos}, which represents the inverse uniform
susceptibility, becomes \pgnformula{fmodemodea}{a\rightarrow
\alpha(p-p_c)+\frac{5}{3}b\left\langle\left|\fatm\right|^2\right\rangle}
\pgnformula{fmodemodeb}{\left\langle\left|\fatm\right|^2\right\rangle=
\sum_{\fatq}\left\langle\left|\fatm_{\fatq}\right|^2\right\rangle}
\pgnformula{fmodemodec}{\left\langle\left|\fatm_{\fatq}\right|^2\right\rangle
=\frac{2}{\pi}\int_0^{\infty}\mbox{d}\omega\left(n_{\omega}\right)Im
{\chi_{\fatq\omega}}} where
$\left\langle\left|\fatm_{\fatq}\right|^2\right\rangle$ is the
thermal variance of a Fourier component of the fluctuating
component of the magnetization and the sum is per unit of volume.
The thermal variance
$\left\langle\left|\fatm_{\fatq}\right|^2\right\rangle$ is defined
by the fluctuation-dissipation theorem in terms of $n_{\omega}$
and the wavevector and frequency dependent susceptibility
$\chi_{\fatq\omega}$. In our model the latter is given by
Equations~\ref{feos}\ and \ref{frelax}, but with $a$ replaced as
in Equation~\ref{fmodemodea}. This yields a self-consistent
equation for
$\left\langle\left|\fatm_{\fatq}\right|^2\right\rangle$ and thus
the magnetic equation of state versus temperature and pressure.
(Note that the zero-point contribution to the total variance of
the local magnetization is not included in
Equation~\ref{fmodemodec}. For further discussions see, e.g.,
Refs.~31 and 43.)

The Curie temperature is defined by a vanishing linear coefficient
in the magnetic equation of state.
Equations~\ref{feos}-\ref{fmodemodec} then yield under our
assumption and for $p<p_c$, \pgnformula{fTCurie}{T_{Curie}\approx
2.39
c\gamma^{\frac{1}{4}}\left(\frac{\alpha(p_c-p)}{b}\right)^{\frac{3}{4}}}
Therefore, near $p_c$, the Curie temperature should vary as $(p_c
- p)^{3/4}$ and depend solely on the ground state parameters of
the model defined by Equations~\ref{feos} and \ref{frelax}. This
result holds only for sufficiently low values of $T_{Curie}$ where
the breakdown region of the mean-field approximation around
$T_{Curie}$ is sufficiently narrow. From the measured parameters
Equation~\ref{fTCurie} yields a value of $T_{Curie}$ of the order
of 40~K for \nal\ at ambient pressure, in good agreement with
experiment.\cite{lon85a}\ Our observed pressure dependence of
$T_{Curie}$ is not inconsistent with the model, although more
detailed measurements near $p_c$ are needed to provide a test of
the predicted $(p_c-p)^{3/4}$ variation of $T_{Curie}$. The
phenomenological parameters in the model might be inferred from an
appropriate energy-band model that includes effects of zero-point
spin fluctuations. Recently, such a model has been considered and
it leads in particular to insights into the origin of the pressure
dependence of the ordered moment in \nal.\cite{agu04a,maz04a}

\subsection{Temperature dependence of the resistivity}

Next we consider the effect of thermal spin fluctuations on the
resistivity. Within the Boltzmann-Born model \cite{zim60a}\ one
finds that $\Delta\rho$ in the paramagnetic state can be expressed
in the form
\pgnformula{frho}{\Delta\rho=\eta\sum_{\fatq}q\left(T\frac
{\partial\left\langle\left|\fatm_{\fatq}\right|^2\right\rangle}
{\partial T}\right)_{\Gamma_q}} where $\eta$ is a constant. The
temperature derivative is solely with respect to the Bose function
entering the definition of the thermal variance. We have assumed
that momentum is efficiently transferred from spin fluctuation to
the lattice either by Umklapp processes or residual disorder. The
thermal variance of the local magnetisation arises in
Equation~\ref{frho} because current carriers are assumed to
undergo weak scattering from a molecular exchange field that is
proportional to the local magnetization. Multiple scattering
processes are not included. The factor $q$ in Equation~\ref{frho}
comes from a product of (i) a $q^2$ factor arising from the fact
that low $q$ fluctuations are ineffective in reducing the current
and (ii) a partly compensating $1/q$ factor due to a loss of
wavevector phase space coming from the Pauli principle that
constrains scattering to the vicinity of the Fermi surface.
Corrections due to momentum non-conservation possibly arising from
residual disorder or intraband transitions are not included. The
temperature derivative $T\partial/\partial T$ can be shown to
arise essentially from the fact that the available momentum phase
space increases with increasing temperature due to the thermal
depopulation of states below the Fermi surface.

\pgnfigure{6}{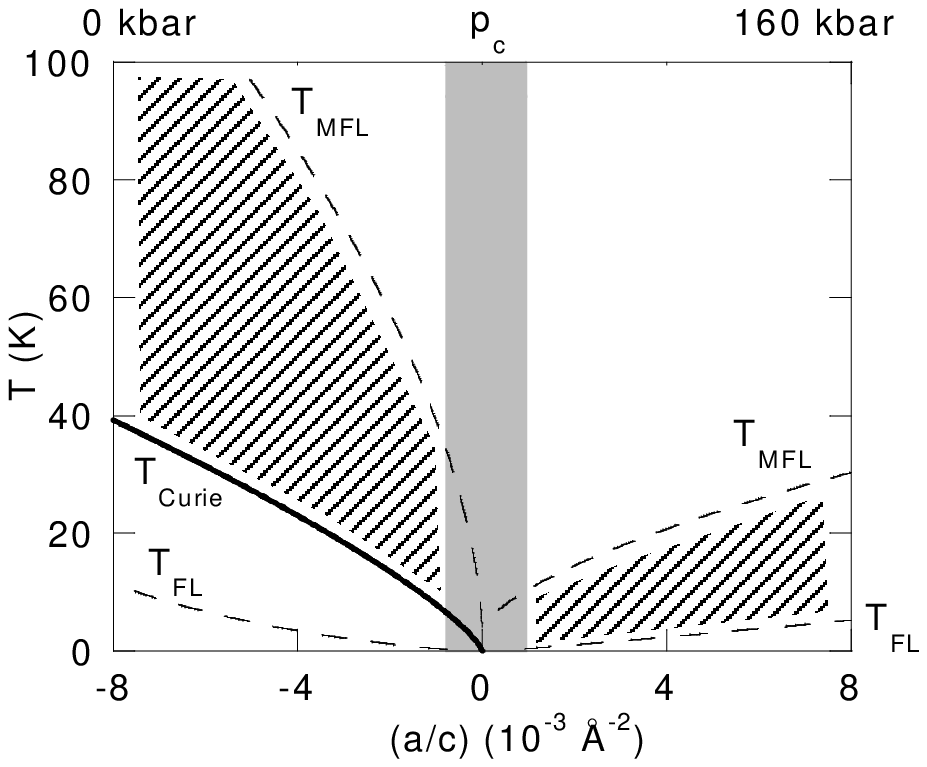}{Temperature-pressure phase diagram
predicted by the mean-field spin-fluctuation model discussed in
the text.  For $p>p_c$ the lower axis represents the square of the
magnetic correlation vector (inverse magnetic correlation length)
in the zero temperature limit. The solid line represents the Curie
temperature given by Equation~\ref{fTCurie}. The crossover lines
$T_{FL}$ and $T_{MFL}$ are defined in the text. If the
ferromagnetic transition becomes first order at low temperatures,
then there exists a region (vertical shading) of forbidden values
of $a/c$.\cite{nik04b}}

Below a characteristic temperature $T_{FL}$ the above model
(Equations~\ref{feos}-\ref{frho}) leads to $\Delta\rho$ of the
form $\Delta\rho=AT^2$ where $A$ is inversely related to the
square of the temperature $T_{FL}$ that vanishes as $p\rightarrow
p_c$. At this pressure $\Delta\rho$ is predicted to vary as
$T^{5/3}$ in the low $T$ limit. An elementary discussion of (i)
the connection of $T_{FL}$ to the model parameters and (ii) the
$T^{5/3}$ form of the resistivity and its connection to the
marginal Fermi liquid model may be found for example in
Ref.~\onlinecite{pfl97a}.

An extension of the above treatment into the ferromagnetic
state\cite{mor85a,lon85a}\ leads again to a $T^2$ resistivity
below a characteristic temperature $T_{FL}$ and a $T^{5/3}$
resistivity above $T_{Curie}$, but below another characteristic
temperature $T_{MFL}$. Above $T_{MFL}$ the resistivity exponent
decreases and tends to unity at higher $T$ due to the growth of
the thermal variance of the local magnetization in
Equation~\ref{fmodemodea}, i.e., due to the coupling between spin
fluctuation modes in the mean field approximation.

\subsection{Temperature-pressure phase diagram for \nal}

The above findings are summarized in the predicted
temperature-pressure phase diagram for \nal\ shown in
Figure~\ref{ppnalphdiagsim}. The pressure dependences of
$T_{Curie}$ below $p_c$ and of $T_{MFL}$ and $T_{FL}$ in the
paramagnetic state below or above $p_c$ are obtained from (i)
Equations~\ref{feos}-\ref{frho}, (ii) the model parameters as
defined in Ref.~\onlinecite{lon85a}, and (iii) $p_c=82$~kbar. The
values of $T_{FL}$ below $p_c$ are obtained by an extension of the
model to the ferromagnetic state.\cite{mor85a,lon85a}\ The
ferromagnetic transition is assumed to be everywhere continuous.
The Curie temperature $T_{Curie}$ is defined by
Equation~\ref{fTCurie}. The characteristic crossover temperature
$T_{FL}$ is defined by the condition
$n(T_{FL})=ln\Delta\rho/lnT=1.8$ and the crossover temperature
$T_{MFL}$ by the condition $n(T_{MFL})=1.6$.

\subsection{Comparison with experiment}

\pgnfigures{6}{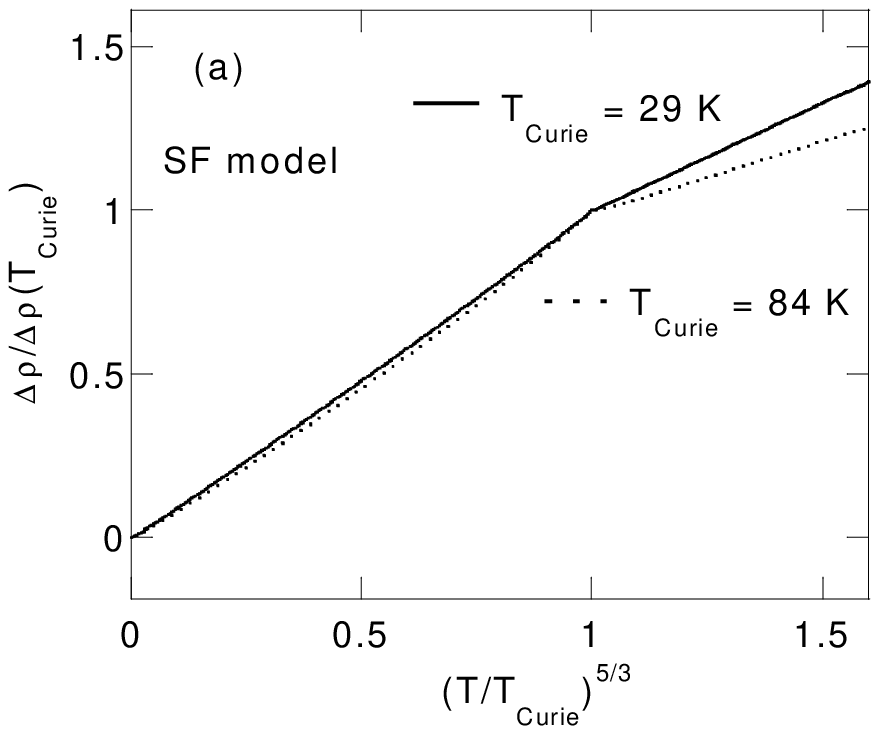}{6}{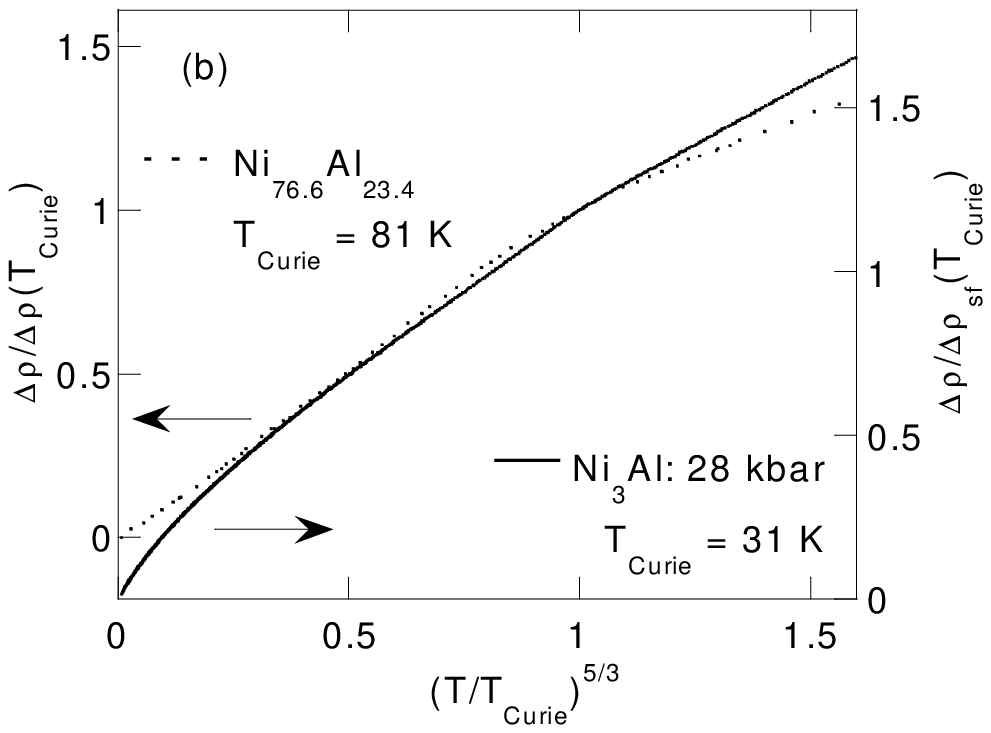}{The temperature
dependent part of the resistivity $\Delta\rho$ vs $T^{5/3}$
normalized to values at $T_{Curie}$. Predictions of the mean-field
spin-fluctuation model (a) and experimental findings (b) for two
samples with different values of $T_{Curie}$. The higher value of
$T_{Curie}$ was obtained by chemical doping\cite{ful92a} and the
lower value by the application of hydrostatic pressure to our
stoichiometric samples (Figure~\ref{ppnaldata}a). The
correspondence between theory and experiment is strikingly close
except for the sample with the lower $T_{Curie}$ at low
temperatures. This discrepancy at low $T$, which indicates an
extra component in $\Delta\rho$ additional to the spin-fluctuation
component $\Delta\rho_{sf}$, extends up to $p_c$ and beyond. It is
also visible in earlier ambient pressure data in a stoichiometric
sample.\cite{sas84a,nik04b,nik04c}}

We find that the experimental results may be understood quite well
in terms of the above model over a wide range in temperature and
pressure, except in the range of about 1-10~K. The regions of
agreement and disagreement between the model and experiment are
illustrated in the examples shown in Figure~\ref{ppnalrft_theo}.
The variation of $\Delta\rho$ vs $T^{5/3}$ in two samples of \nal\
with different $T_{Curie}$ is plotted in normalized form in
Figure~\ref{ppnalrft_theo}b.\cite{ful92a}\ The predictions of the
model for the set of parameters appropriate to the two samples are
shown in Figure~\ref{ppnalrft_theo}a.\cite{lon85a}\ The agreement
between theory and experiment is rather striking both in the
magnitudes of $T_{Curie}$ and the overall form of the curves.

However, the stoichiometric sample at high pressure exhibits an
anomalous downturn at low $T$ which is not anticipated by the
model. This discrepancy is present in all of our high precision
measurements in stoichiometric samples up to $p_c$ and beyond. It
is also evident in an earlier study at ambient
pressure.\cite{sas84a}\ The above model predicts a $T^{5/3}$ form
of $\Delta\rho$ at all $T$ below approximately 10~K at the
critical pressure. This is not observed (Figure~\ref{ppnaldexpT}).
Instead, as discussed earlier, we observe a continuous change in
$x(T)$ varying from approximately 1.3 towards 2 as the temperature
is decreased from 5~K to below 1~K.

The $T^2$ temperature dependence seen away from $p_c$ is
qualitatively consistent with the model. However, the magnitude of
$T_{FL}$ estimated from Figure~\ref{ppnaldexpT}\ or from
Figure~\ref{ppnaldsqdev}\ appears to be nearly an order of
magnitude lower than predicted by the model
(Figure~\ref{ppnalphdiagsim}).

\subsection{Corrections to the elementary mean-field
spin-fluctuation model}

The existence of a $T^2$ regime at $p_c$ well below 1~K is not
expected if the ferromagnetic transition is second order. A finite
$T_{FL}$ at $p_c$ may indicate that the phase transition becomes
first order close to $p_c$ as might be generally expected for
ferromagnetic quantum phase transitions.\cite{bel99a,chu04a}. Here
we note that, as found in MnSi, a first order transition need not
lead to a Fermi liquid form of $\Delta\rho$ near $p_c$ (i.e.,
where $T_{Curie}$ vanishes).\cite{doi03b}

The most significant challenge to the mean-field spin-fluctuation
model is to account for the anomalous downturn seen in
Figure~\ref{ppnalrft_theo}b, which was not anticipated and remains
unexplained. We stress, however, that in other respects the
mean-field spin-fluctuation model provides a rather accurate
description of the temperature-pressure phase diagram of \nal.

The low temperature form of the resistivity is also poorly
understood in a number of other materials.  In MnSi one observes a
$T^{3/2}$ variation of $\Delta\rho$ over nearly 3 decades below
10~K in an extended range in pressure beyond the critical pressure
of a first order magnetic transition.\cite{pfl01b,doi03b,pfl04a}\
This is at odds with the prediction that $T_{FL}$ should be of the
order of 5~K or higher at all pressures in this system. In
$\epsilon$-Fe the resistivity varies as $T^{5/3}$ over two decades
above the critical pressure of a first order magnetic and
structural phase transition. Naively one would have expect to see
a $T^2$ resistivity below a few K in this system
too.\cite{jac02a,jar02a}

Non-Fermi liquid forms of the resistivity have now been observed
in a large number of examples in which the applicability of the
mean-field spin-fluctuation model is questionable (see, e.g.,
Refs~\onlinecite{ste03a,cap02a,riv03a}).  The cases of MnSi and
\nal\ may stand out partly because the magnitudes of the
discrepancies between the predictions and the experimental
findings can be quantified in terms of independently determined
model parameters. Also, in these two materials the discrepancies
arise in small pockets of the temperature-pressure phase diagrams
surrounded by wide regions where the mean-field spin-fluctuation
model provides a rather accurate description of the experimental
findings.

The systems in which the discrepancies between theory and
experiment are most clearly evident are characterized by first
order transitions for sufficiently low $T_{Curie}$.  This implies
that the parameter $b$ in Equation~\ref{feos}\ must be taken to be
negative, i.e., the coupling between the spin-fluctuation modes is
attractive. The applicability in this case of the mean-field
decoupling method is not self-evident. Systems described by a
quantum Ginzburg-Landau model in which  $b<0$ (but higher order
coefficients in the expansion in M are positive) may be
characterized by highly anharmonic magnetic fluctuations not
described in the mean-field approximation. Such fluctuations may
be of particular interest in cases when the tricritical point
characteristic of this extended quantum Ginzburg-Landau model is
at sufficiently low temperatures.

Beyond this one may also look for an explanation of the
discrepancies between experiment and the mean-field
spin-fluctuation model in a number of more conventional effects
arising from lattice vibrations,\cite{flu69a,sta81a}\ band
degeneracies,\cite{mor81a}\ Umklapp processes\cite{ash76a}\ and
quenched disorder. A systematic study of such effects is still
lacking for the systems we have considered. We note, however, that
the effects of conventional phonon scattering, in particular, is
expected to be very weak and ignorable in \nal\ at least below
100~K.\cite{flu69a}\ Also, the effects of quenched disorder may be
expected to be strongest below 1~K where $\Delta\rho\ll\rho_0$ in
our \nal\ samples. However, the behaviour of $\Delta\rho$ vs $T$
in this regime is not inconsistent with our model predictions
except perhaps very near $p_c$.  For example, $\Delta\rho/T^2$ at
500~mK is predicted to increase by a factor of about 2 from
ambient pressure to $p_c$. This agrees well with the variation of
$A$ shown in the inset of Figure~\ref{ppnaldlt}b. These
considerations suggest that the anomalous behaviour of
$\Delta\rho$ in the range 1-10~K may arise from essentially
intrinsic effects of the electron-electron interaction and not
explicitly from phonons or quenched disorder.

\section{conclusion}

The magnetic temperature-pressure phase diagram for \nal\ has been
explored for the first time beyond the low-pressure regime by
means of 4 terminal resistivity measurements in a diamond anvil
cell. Hydrostatic pressure studies up to 100~kbar and down to
50~mK indicate that the Curie temperature collapses towards
absolute zero at a critical pressure $p_c=82\pm 2$~kbar.

A non-Fermi liquid form of the resistivity is observed over a wide
range in temperature and pressure except below a characteristic
temperature $T_{FL}$ that decreases from a few Kelvin at ambient
pressure to well below one Kelvin at $p_c$.  The finite value of
$T_{FL}$ at $p_c$ may indicate that the ferromagnetic transition
is first order near the critical pressure. The temperature
dependence of the resistivity and $T_{Curie}$ is, overall,
consistent with the predictions of the mean-field spin-fluctuation
model, which reduces to a type of marginal Fermi liquid model just
above $T_{Curie}$ when $T_{Curie}$ tends towards absolute zero.
However, a comparison of theory and experiment also reveals the
existence of an extra component in the temperature dependence of
the resistivity at intermediate temperatures that has a different
and unidentified origin.


\begin{acknowledgments}

We wish to thank  S. Julian, A. Rosch, A. Huxley and S. Yates for
stimulating discussions on this topic. PGN is grateful for support
by the FERLIN program of the European Science Foundation.

\end{acknowledgments}


\begin{thebibliography}{68}
\bibitem[{nik({\natexlab{a}})}]{nik04a}
\bibinfo{note}{A relaxation rate linear in $E$ might be expected to give rise
  to a resistivity varying as $T$ rather than as $T^{5/3}$. The extra factor of
  $T^{2/3}$ arises in our case because spin-fluctuations or paramagnons on the
  border of ferromagnetism give rise to scattering mainly in the forward
  direction and are thus relatively ineffective in reducing the current.}
\bibitem[{nik({\natexlab{b}})}]{nik04b}
\bibinfo{note}{The parameters of our model for stoichiometric \nal\ at ambient
  pressure are given in Ref.~\onlinecite{lon85a}. It is assumed that only the
  critical parameter $a$ varies appreciably with pressure $p$ or doping $x$.
  The value of $a$ as a function of $p$ and $x$ is obtained from its value at
  $p=x=0$ times $\beta(p,x)=M_0^2(p,x)/M_0^2(0,0)$, where $M_0(p,x)$ is the
  magnetization in the zero temperature and zero field limit (see
  Equation~\ref{feos}). In the analyses presented in this paper, $\beta(0,x)$
  is inferred from the data in Ref.~\onlinecite{ful92a}\ and $\beta(p,0)$ is
  taken to be $(1-p/p_c)$ for $p<p_c$ (see Equation~\ref{feos}\ and the assumed
  form of $a$ vs $p$).}
\bibitem[{nik({\natexlab{c}})}]{nik04c}
\bibinfo{note}{The data for the sample with the lower $T_{Curie}$ is only
  normalised with respect to $\Delta\rho_{sf}(T_{Curie})$. Motivated by the
  theoretical simulation $\Delta\rho_{sf}(T_{Curie})$ is determined by a
  $T^{5/3}$ fit to the data from 10~K to $T_{Curie}$ and extrapolation of this
  fit to $T=0$~K.}
\end{thebibliography}

\end{document}